\documentclass[preprint,12pt]{elsarticle}

\usepackage{amssymb}
\usepackage{amsmath}

\newcommand\sref[1]{Sec.~\ref{#1}}
\newcommand\fref[1]{Fig.~\ref{#1}}

\newcommand{\eg}{\emph{e.g.}}
\newcommand{\ie}{\emph{i.e.}}

\newcommand\unit[2]{#1\,#2}
\newcommand\unitr[2]{\unit{#1}{\mathrm{#2}}}
\newcommand\Ang{\text\AA}

\begin{document}
\title{All-graphene edge contacts: Electrical resistance of graphene T-junctions}
\author[cng,pho]{K{\aa}re Wedel Jacobsen}
\author[cng,nano]{Jesper Toft Falkenberg}
\author[cng,nano]{Nick Papior}
\author[cng,nano]{Peter B{\o}ggild}
\author[cng,nano]{Antti-Pekka Jauho}
\author[cng,nano]{Mads Brandbyge\corref{cor1}}
\cortext[cor1]{Corresponding author. Tel: +45 45 25 63 28. E-mail: mads.brandbyge@nanotech.dtu.dk (Mads Brandbyge)}
\address[cng]{Center for Nanostructured Graphene (CNG)}
\address[pho]{Dept. of Photonics Engineering, Technical University of Denmark, {\O}rsteds Plads, Bldg.~343, DK-2800 Kongens Lyngby, Denmark}
\address[nano]{Dept. of Micro- and Nanotechnology, Technical University of Denmark, {\O}rsteds Plads, Bldg.~345E, DK-2800 Kongens Lyngby, Denmark}

\begin{abstract}
  Using \emph{ab-initio} methods we investigate the possibility of three-terminal graphene
  ``T-junction'' devices and show that these all-graphene edge contacts are energetically
  feasible when the 1D interface itself is free from foreign atoms. We examine the
  energetics of various junction structures as a function of the atomic scale
  geometry. Three-terminal equilibrium Green's functions are used to determine the
  transmission spectrum and contact resistance of the system. We find that the most
  symmetric structures have a significant binding energy, and we determine the contact
  resistances in the junction to be in the range of $1-10~{\rm k}\Omega\,\mu{\rm m}$ which
  is comparable to the best contact resistance reported for edge-contacted graphene-metal
  contacts\cite{Akinwande2014,Wang2013}. We conclude that conducting all-carbon
  T-junctions should be feasible.
\end{abstract}

\maketitle

\section{Introduction}
\label{sec:intro}

Two-dimensional (2D) materials are being vigorously investigated as a platform for nano-scale electronics due to their potential use \eg\ in flexible electrodes\cite{Akinwande2014}, high performance electronics, photovoltaics and spintronics\cite{Novoselov2012}. Graphene plays a key role, not only because it was the first 2D material to be isolated and experimentally characterized\cite{Geim2007}, but also because of its extraordinary electronic properties, which can be harnessed by various types of nanostructuring and chemical functionalization\cite{Han2007,pedersen_graphene_2008,Elias2009,Ho2014,Sofo2007,Oostinga2008,Ohta2006,Rasmussen2013}. 

\begin{figure}[t]
  \centering
  \includegraphics[width=0.5\columnwidth]{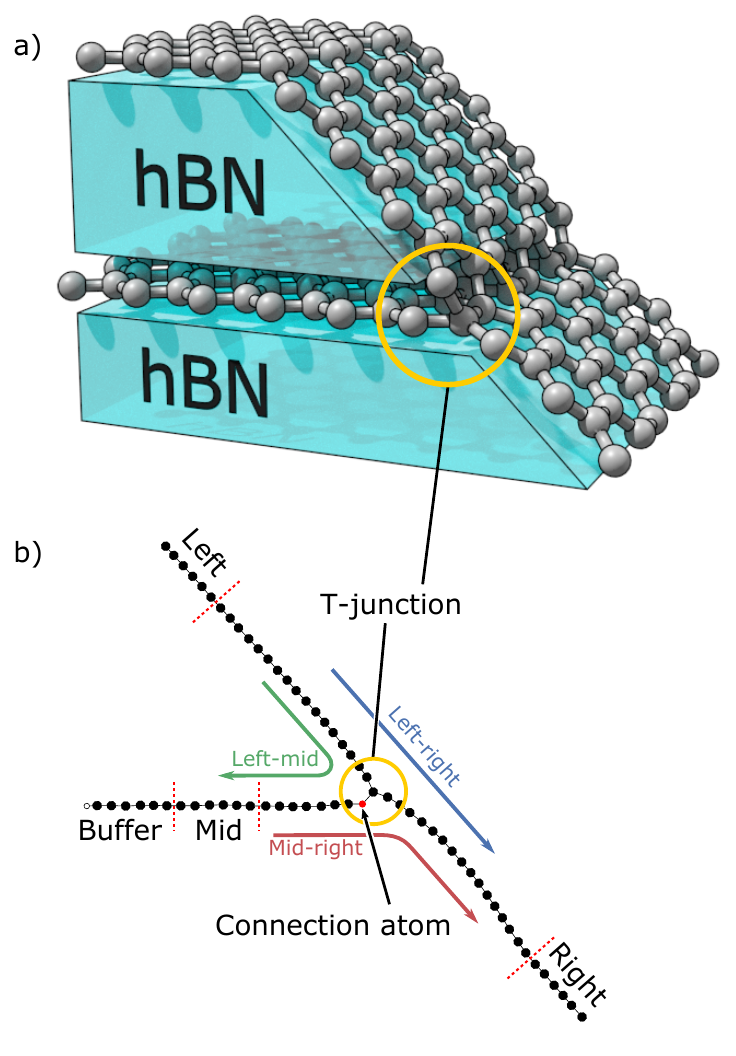}
  \caption{\textbf{a)} Sketch of a possible experimental realization where the GTJ is partly encapsulated in a stack of hexagonal boron-nitride. \textbf{b)} Schematic of the graphene T-junction system (side view) consisting of 3 semi-infinite graphene electrodes (``Mid''/``Buffer'', ``Left'', ``Right'') used in the DFT-EGF calculation. In the calculations that include passivation, the connection atom was substituted with other elements. The structure shown here is the result of a nitrogen substitution in the ZZ geometry.
      \label{fig:setup}
    }
\end{figure}

Recently graphene has played a role as contact electrode to semi-conducting transition metal di-chalcogenides (TMDC) encapsulated in insulating hexagonal boron-nitride (hBN) layers\cite{Wang2013}. In these devices graphene is connected to the external circuit via one-dimensional edge-contacts to 3D metal electrodes\cite{Wang2013,Cui2015}. In the fabrication process the hBN-G-hBN stack is etched with a slope of approximately $45^\circ$ resulting in a graphene edge being exposed to subsequent metal electrode deposition.

While these studies have so far focused on planar devices it is relevant to investigate various ways of extending the 2D structures into 3D circuitry. To this end, and inspired by the experimentally realized 1D edge contacts\cite{Wang2013}, we here use first principles calculations to investigate graphene T-junctions (GTJ), where the bulk metal electrode is replaced by graphene as illustrated in \fref{fig:setup} and \ref{fig:structs}. The out-of-plane bonding is possible due to the sp$^2$ nature of graphene, which can hybridize further to sp$^3$ and thus allows the formation of a ``standing'' sheet (or ribbon). The electronic properties of the graphene T-junctions are only limited by the junction itself as the long range ripples are an intrinsic detail in graphene\cite{Fasolino2007}.
Formation of such a T-junction requires the edge atoms of one layer to form covalent bonds to the plane of another layer, which can either be done by fusing or by synthesis. Coalescing or fusing of separate carbon nanostructures can either be achieved through Joule heating\cite{qi2015electronic,zou2012method}, ion\cite{wu2014molecular,wu2013formation} or electron\cite{terrones2000coalescence} irradiation, where the extraordinary ability of sp$^2$ carbon nanostructures  to self-repair\cite{zan2012graphene} can be exploited to reach well-defined, stable, low-energy configurations.
In principle, the alternative approach of bottom-up synthesis can can also lead to creation of complex hybrid all-carbon architectures with interconnections such as graphene carbon-nanotubes\cite{Akinwande2014,Yan2014}, 3D interconnected graphene ``foam''\cite{Chen2011b} and vertical T-junction-like ``nanowalls''\cite{Lisi2011,Kumar2014}.
While graphene-nanotube two-terminal systems have been investigated by first principles calculations\cite{Novaes2010}, the electronic transport properties of T-junctions involving three semi-infinite graphene or graphene nano-ribbon (GNR) electrodes have not, to the best of our knowledge, been investigated. 
The paper is organized as follows: In \sref{sec:sysmeth} we describe the systems and computational method. In \sref{sec:results} we present the results for the energetics and structure of both infinitely wide junctions and narrow ribbon junctions to the graphene plane, and discuss the electronic transmission as a function of electronic energy (doping level or gate voltage) and electrical contact resistance. Finally, we summarize the results and present an outlook of future work and experimental realization of graphene T-junctions in \sref{sec:conclusions}.

\section{Systems and methods}
\label{sec:sysmeth}

\begin{figure}[htb]
	\centering
	\includegraphics[width=.3\columnwidth]{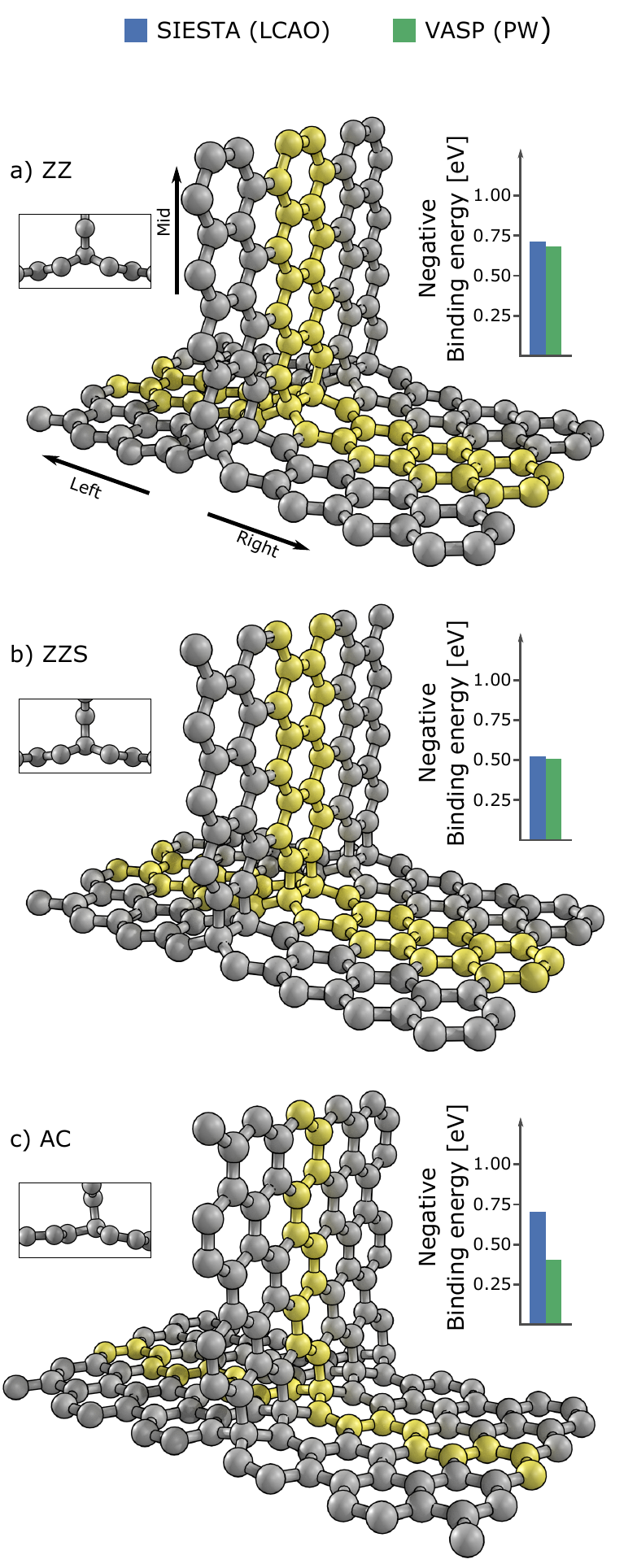}
	\caption{Sketch showing the three main principal structures of interest. The structures are periodic in the transverse direction and yellow indicates atoms belonging to the unit cell. The insets shows the binding energy of the middle part as calculated with the SIESTA~(blue bar) and VASP~(green bar) codes which are in reasonable agreement. The insets show sideviews of the structures, clearly illustrating the sp$^3$ geometry at the junction.}
	\label{fig:structs}
\end{figure}

In order to predict the atomic structures and their binding energy we employ density
functional theory calculations (DFT). Electronic transport is calculated by the Green's function method\cite{Brandbyge2002} extended to the
multi-terminal case, which in the present work involve three electrodes denoted ``Left'',
``Right'' and ``Mid'', see \fref{fig:setup}. The middle electrode corresponds to a
graphene layer which for instance could be encapsulated in a hBN stack as shown in
\fref{fig:setup}. Here only non-gated graphene electrodes are considered, while in general
care must be taken to account properly for gating effects of graphene
electrodes\cite{Papior2016}. The geometrical structure of the junction is, as we shall
see, a main factor determining the resistance.
 
We first focus on the two main symmetry directions of current flow in graphene (zig-zag and armchair) and consider three different principal structures corresponding to perfect match in the junction as shown in \fref{fig:structs}. We note that the structures have been rotated compared to \fref{fig:setup} so that the ``mid''-section is pointing out-of-plane of the graphene sheet going from ``left'' to ``right''. The simplest possible connection is denoted ``ZZ'' according to the zig-zag edge of the attached ``ribbon'' as shown in \fref{fig:structs}a, and is mirror symmetric around the middle part and involves only hexagonal carbon rings. The second principal structure is a shifted zig-zag (ZZS) and is also mirror symmetric, but the middle part has been shifted one half of a unit cell resulting in 4- or 8-ring transition in the T-junction. The armchair junction (AC) in \fref{fig:structs}c has no mirror-symmetry between the electrodes. However, it is noted that the left-mid and the mid-right transitions have similar grain boundary types in the junctions. In \sref{sec:transmission} we will see how this similarity is reflected in the transmission through the junction. For all these structures we employ periodic boundary conditions (PBC) along the one-dimensional junction. We have also investigated junctions with an initial angle different from $90^\circ$ between the ``mid'' section and the base sheet. When allowed to move freely, however, the ``mid'' section relaxes towards the symmetric configuration in all cases.

\begin{figure*}[t]
  \centering
  \includegraphics[width=.9\linewidth]{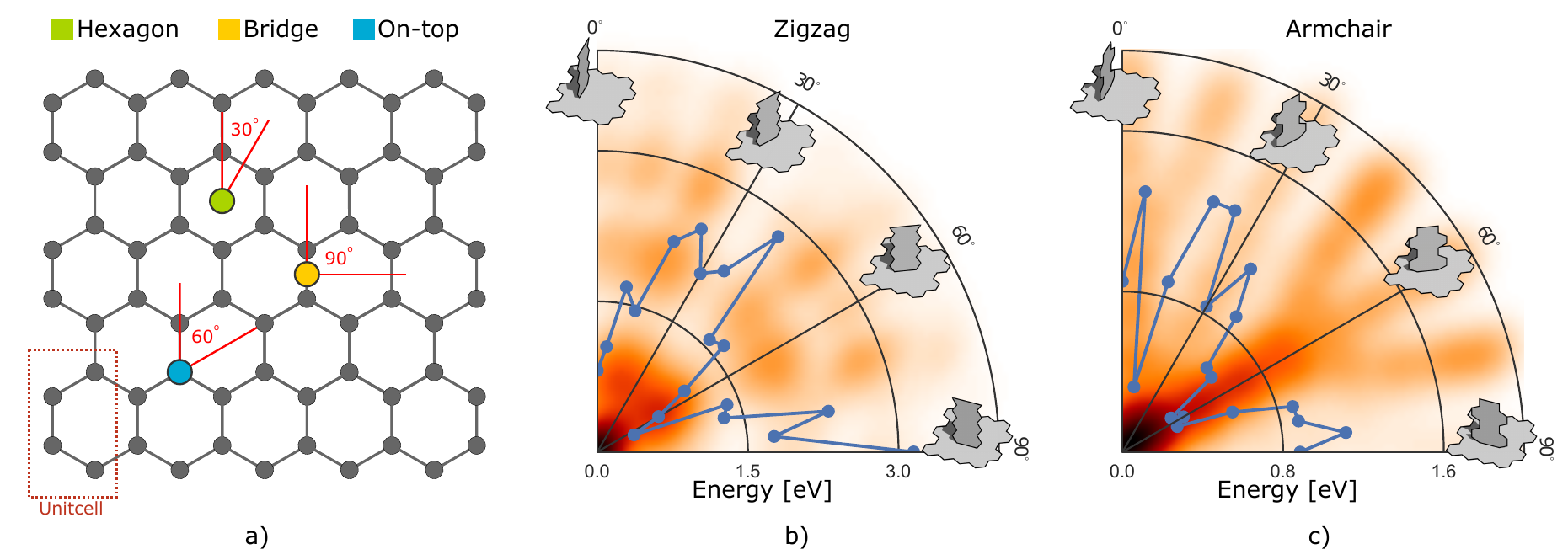}
  \caption{\textbf{a)} The setup for the starting configurations of the calculations of ribbon T-junctions. Each calculation was initiated with the
      ribbons placed one bondlength above the base sheet and with the bottom corner atom
      above one of the three sites, hexagon (green), bridge (yellow) or on-top (blue). The
      ribbons were rotated around the anchor points in steps of $5^\circ$ between the
      directions indicated with red.
      \textbf{b)}, \textbf{c)} The blue curve shows the mean energy of the three ribbons started at each angle
      (one for each site). It is shown atop a gaussian kernel density plot (red colors) of the two bottom rows of atoms
      in each of the relaxed ribbons translated to have the outermost atom in the same
      point and projected onto the plane. The orientation of the underlying graphene sheet is as in a). Both types of ribbons energetically favor the $0^\circ$ and $60^\circ$ directions, making the zig-zag ribbon similar to the ZZ(S) structure in \fref{fig:structs}, but the armchair ribbon different from AC in the same figure.
      \label{fig:ribbons}
  }
\end{figure*}

In order to examine the role of a less symmetric match in the T-junction we also investigate \emph{rotated} junctions by attaching the nano-ribbons in the ``Mid'' position on top of an infinite graphene sheet (still containing ``Left''/``Right'' electrodes). Both zig-zag and armchair nano-ribbons are investigated by placing the bottom right carbon atom one bond-length above the graphene sheet consisting of $10\times 5$ 4-atom unit cells (indicated in \fref{fig:ribbons}). The attachment site on the graphene sheet is chosen among the three high-symmetry sites, \emph{on-top}, \emph{bridge}, and \emph{hexagon}, as illustrated in \fref{fig:ribbons}a. For each attachment site, the ribbons are rotated in steps of five degrees within the angles as indicated in the figure.

In summary, with the use of periodic boundary conditions, we examine T-junctions formed by infinte graphene sheets as shown in \fref{fig:structs} and finite-width ribbons intersecting with an infinite graphene layer as depicted in the small illustrations in \fref{fig:ribbons}.

\subsection{Computational methods}
\label{ssec:theory}

The DFT calculations were performed using the software packages SIESTA\cite{Soler} and
VASP\cite{kresse1996efficient}. The former utilizes a localized basis set (LCAO), which
allows much faster calculations than the the more accurate plane-wave basis in VASP, which
provides a quality check on the total energies of the LCAO method. A SZP basis set was
chosen for SIESTA after noting only negligible differences in the resulting relaxed
geometries and transmission spectra when comparing to a DZP basis set. The plane-wave
calculations used a cut-off of $\unitr{400}{eV}$.  The PBE-GGA functional for
exchange-correlation\cite{Perdew1996} was used for both methods, as well as an atomic
force tolerance of $\unitr{0.04}{eV/\Ang}$. Additionally, the SIESTA calculations used
confinement radii determined from an energy shift of $\unitr{275}{meV}$ with a real-space
grid cutoff energy of $\unitr{300}{Ry}$. The PBC of the principal structures along the 1D
junction was utilized in $k$-point sampling by a Monkhorst-Pack grid of $15 \times 1 \times
1$ ensuring relative energy convergence. In the subsequent transport calculations, the three electrodes (``left'', ``right'', and ``mid'') were all treated as semi-infinite while the system was modelled as periodic in the transverse direction. The transport calculations were
performed using the TranSIESTA\cite{Brandbyge2002,Papior2016a} method extended with a
recently implemented $N$-electrode capability following \citet{Saha2009}. This allows the
description of proper boundaries for the three semi-infinite graphene leads. All
electrodes are described using surface self-energies from separate bulk calculations. Here we focus on low bias properties and 
only present equilibrium transport calculations while full non-equilibrium calculations are presented
elsewhere\cite{Papior2016a}. We extended the ``Mid'' electrode using ``buffer''
atoms\cite{Brandbyge2002} in order to obtain a bulk electrode potential profile for
this. Transmission calculations are performed using 100 $k$-points and post-processed using
the interpolation technique described by \citet{Falkenberg2015} in order to obtain well
converged smooth transmission functions.

\section{Results}
\label{sec:results}
\subsection{Energetics}
\label{sec:energies}

All three structures shown in \fref{fig:structs} have a negative binding energy, and could thus be experimentally feasible. The total energy calculations were consistent when comparing energies from SIESTA with VASP after relaxing the atoms, as is shown as bars to the right of the structures in \fref{fig:structs}.

We investigate the rotation of ribbons perpendicularly attached to a graphene sheet, but attached to the sites shown in \fref{fig:ribbons}a and allowing the atoms to relax. In the calculation, one row of atoms in the base sheet was fixed in space while all other atoms -- including the entire ribbon -- was free to relax. The zig-zag and armchair ribbons were both four rows of atoms wide. Only negligible rotation is observed of the carbon atoms in the GNRs the furthest away from the base sheet, thus allowing us to define a starting angle. The averaged energy of the three ribbon configurations is shown in \fref{fig:ribbons}b,c. The blue line is the mean energy for the three relaxed structures at each starting angle subtracted the minimum energy configuration of the entire set of structures. The angles in Fig 3b,c indicate the initial rather than the final angle of the relaxed structure, which may be slightly different. In order to examine the relaxed direction a density map of atoms is shown as the background of the energy plot. All relaxed configurations have been translated and projected to the graphene plane. 

We see that the zig-zag ribbons preferentially will be oriented in the directions similar to the principal periodic structure in \fref{fig:structs}a, and are thus corresponding to a geometric transition similar to that of pristine graphene. Note that the armchair ribbons also have the lowest energies when oriented along the armchair direction of the base sheet and are thus {\em not} similar to the periodic armchair structure described earlier, cf. \fref{fig:structs}c. In this configuration, the ribbons are situated symmetrically, but in a way that cannot be periodic without inducing a substantial strain in the graphene layers. \fref{fig:ribbons}c seem to break the $60^\circ$-rotation symmetry as the energies at $0^\circ$ and $5^\circ$ are much larger than the ones around the other armchair direction. This is caused by one of the three calculated structures in each case that quickly relaxed in to a local minimum with very high energy, thus raising the average energy considerably.

\subsection{Transmission}
\label{sec:transmission}

\begin{figure}
  \centering
  \includegraphics[width=.5\columnwidth]{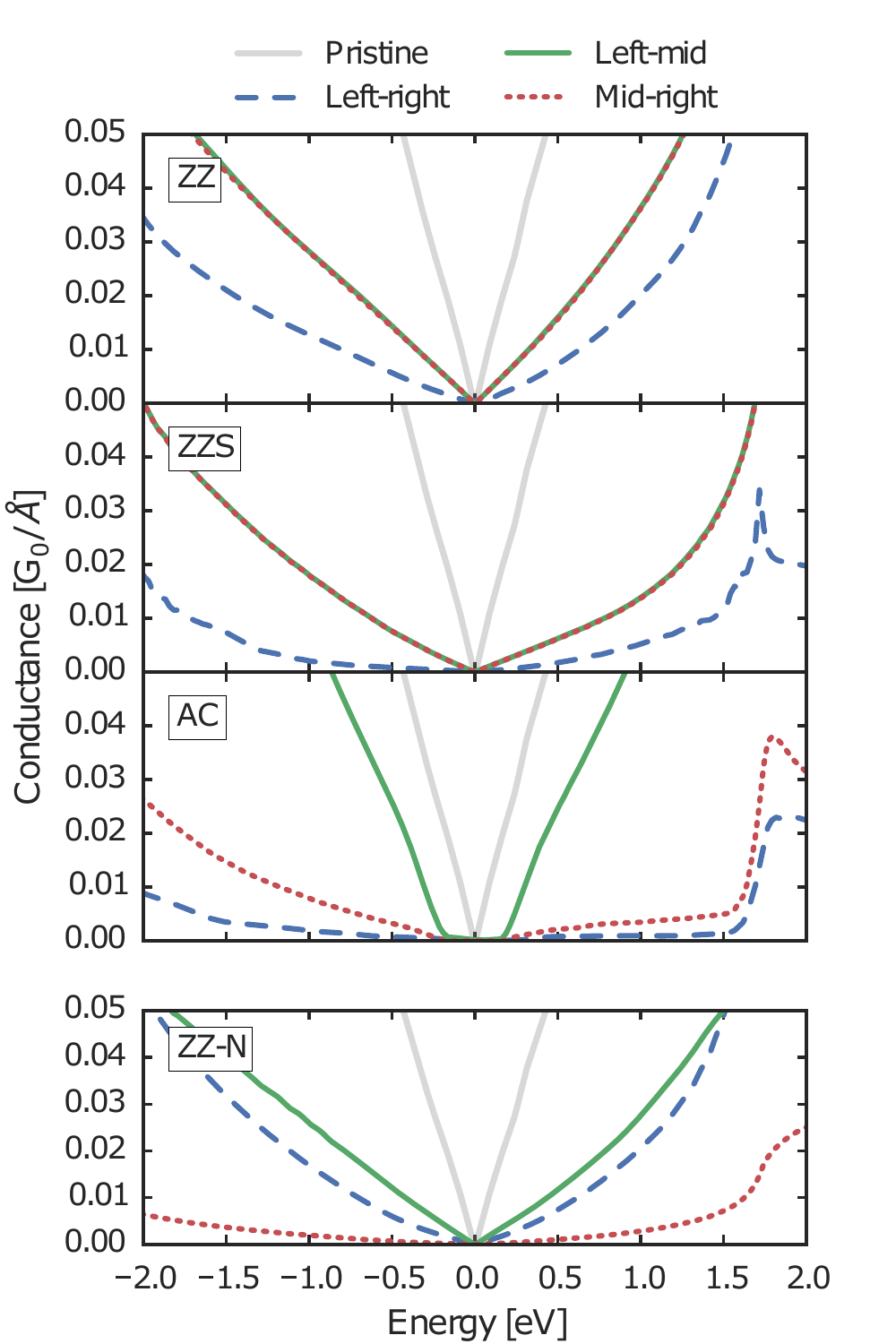}
  \caption{Transmission through the T-junctions for the structures in \fref{fig:structs} (Fermi energy, $E_F=0$). The bottom frame shows the transmission spectrum for ZZ-N where nitrogen replaces carbon as connecting atoms in the arm of the ZZ junction (cf. red atom in Fig. 1b). 
      \label{fig:transmission}
  }
\end{figure}

The transmission spectra shown in \fref{fig:transmission} show conductance per junction length for each of the three principal systems and between each of the three electrodes defined in \fref{fig:structs}. The gray transmission curve is that of pristine graphene. The geometric symmetries of the periodic zig-zag structures (ZZ, ZZS) are reflected in the transmission spectra as the left-mid (blue) and the mid-right (green) curves are identical. Interestingly, it is found that the transmission $T$ through the base sheet (\ie\ left-to-right) is lower than that from the base and into the mid-terminal for all principal structures. In contrast to the ZZ and ZZS structures the armchair (AC) transmission spectrum display a bandgap-like feature for low energies, but yield the highest transmission into the mid-electrode from left (green curve in \fref{fig:transmission}~AC) of all the structures.  

We can compare the left-right transmission $T_\text{left-right}$ through the ZZ structure to that found for hydrogenated kinked graphene\cite{Rasmussen2013}. This can be done by realizing that the row of atoms in the mid-part closest to the junction can be exchanged by hydrogen atoms while the rest of the mid-atoms can be removed. Since the junction atoms are allowed to relax we locally have an sp$^3$-configuration directly comparable to the hydrogenated kinks acting as transmission barriers. There are minor numerical differences in normalized transmission but the trends remain the same: hole transport slowly grows to around $0.03~\text{G}_0/\Ang$ at $-2~\text{eV}$, while the electron transmission grows more rapidly to $0.1~\text{G}_0/\Ang$ at $2~\text{eV}$.

As previously stated, there are similarities in the atomic transitions from left-to-right and from mid-to right in the AC structure. One could thus expect $T_\text{left-right}$ and $T_\text{mid-right}$ to differ from $T_\text{left-mid}$ and this is also what is found in \fref{fig:transmission} AC. The $T_\text{left-right}$ and $T_\text{mid-right}$ transmission spectra are rather similar even though the angle the electrons have to pass through differs greatly.

For the sake of completeness, we have also briefly investigated the effect of forcing the ``mid'' part into an non-orthogonal angle with the base graphene. As previously mentioned, the ZZ(S) T-junctions relax towards the symmetric structures shown in \fref{fig:structs} and to avoid this, we have in one case fixed the ``mid'' terminal part to move doing these relaxations. With an imposed angle of $55^\circ$ between the ``mid'' and ``right'' terminals in an ZZ configuration, only minor changes in the transmission spectrum was observed. The left-to-right and left-to-mid transmissions remained almost unaffected while the mid-to-right lowered a bit to have roughly the same behavior as left-to-right.

\subsection{Passivation}
\label{sub:passivation}

\citeauthor{Wang2013} suggest\cite{Wang2013} that graphene edges exposed after etching might be passivated with H, F or O. They further treat the edges with an O$_2$ plasma to change chemical composition\cite{Felten2005,Bittencourt2011}. After such processing the edges may be oxygen terminated while other contaminants remaining from the edges should have been removed. To investigate the effects of this edge termination on the transport through the junction we  substituted the edge atoms (see connecting atom in \fref{fig:setup}b) of the principal T-junction structures with H, F, O or N. We found, as expected, that H and F passivated edges did not bind to the graphene base sheet. On the other hand we found that while O only bonds in one case (ZZ), N binds to both in the ZZ and ZZS structures. A negative formation energy has been obtained for N on a clean zig-zag edge, originating from a $N_2$ gas\cite{Jiang2012} making incorporation during hBN etching a possibility. Most importantly, both O and N in the connection introduce an angle between the mid part and the base sheet around $45^\circ$, see \fref{fig:setup}b which shows the relaxed ZZ-N structure. The same size as the angle in the hypothetical structure based on the hBN-G-hBN setup due to the etch\cite{Wang2013} as illustrated in \fref{fig:setup}. As seen from \fref{fig:transmission}~(ZZ-N) the introduction of N in the junction and the resulting angle breaks the left-mid and right-mid symmetry in the transmission spectrum. Now left-mid transmission is significantly higher than that of right-mid while the left-right transmission is nearly unchanged compared to the C-only junction (ZZ). Even though the symmetry breaking due to the nitrogen completely changes the transmission, the magnitude of the transmission does not to vary much between the various configurations. We note that similar high left-mid/left-right transmissions were obtained for O passivation.

For all the relaxed structures with a negative binding energy, similar calculations, \ie\ the same number of atoms, were made, but without a connection between the mid part and the base sheet. That is, the atoms were lifted off the graphene sheet such that no contact was possible between the two. This setup was then relaxed, and the energy was used as reference for the binding energy. None of the edge-passivated structures had lower energies when forming a T-junction and are thus unlikely to occur.

\subsection{Junction resistance}

\begin{figure}
  \centering
  \includegraphics[width=.6\columnwidth]{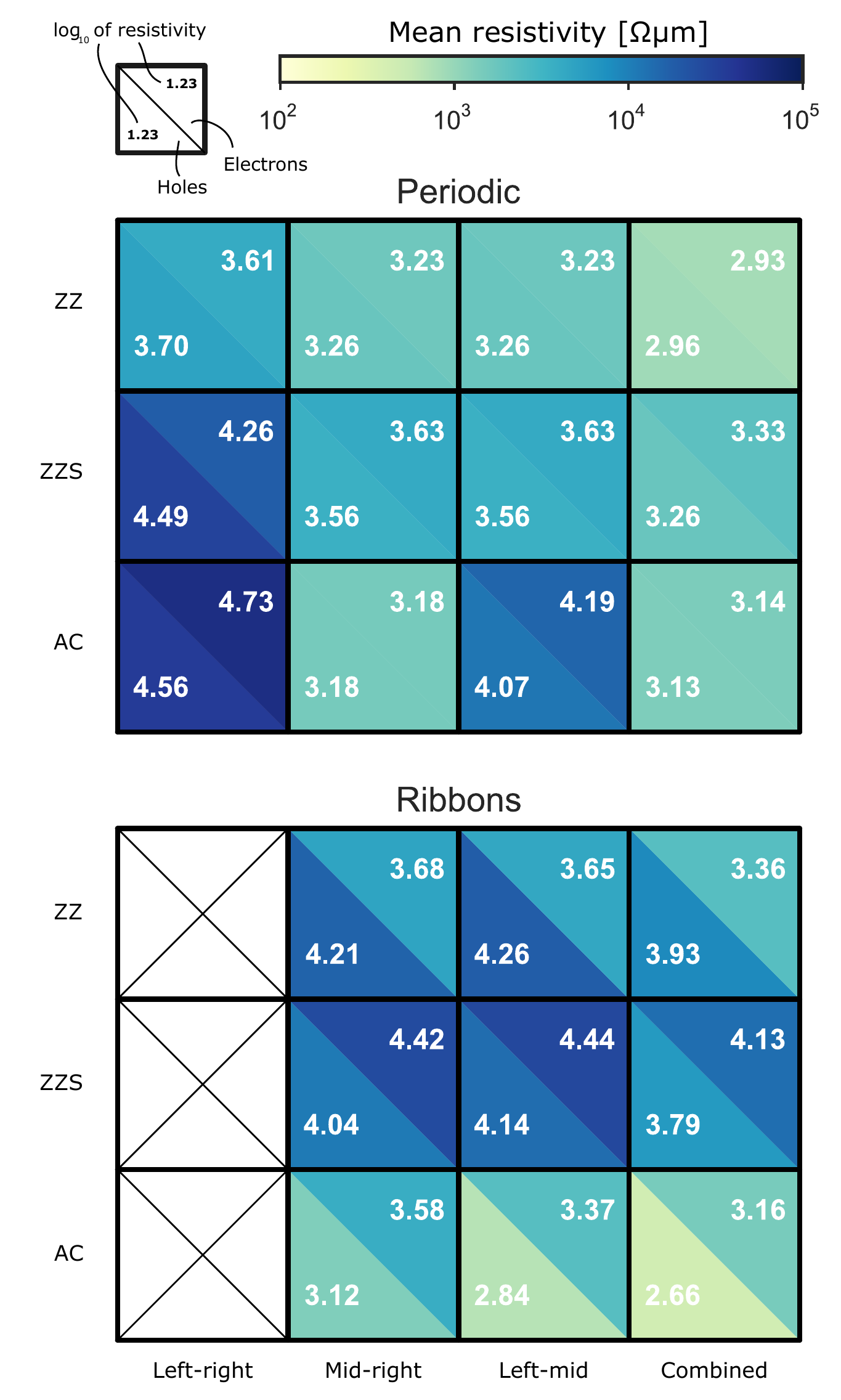}
  \caption{Resistance times length in the T-junctions compared to similar ribbons. The resistivity is taken as the mean 
  	value of electrons with energies between $0\,$eV and $0.5\,$eV on both the hole and the electron side.
      \label{fig:resistance}
  }
\end{figure}

In order to determine the resistance of the contact between the graphene sheet and the attached ``Mid'' graphene terminal, we calculated the mean junction resistance for electrons with energies on each side of the equilibrium Fermi energy ($E_F=0$) in the $\unitr{0}{eV} - \unitr{0.5}{eV}$ range, and likewise for holes, in the $\unitr{-0.5}{eV} - \unitr{0}{eV}$ range. This can be compared to the pristine graphene result for both principal structures and their ribbon counterparts. The result is shown in \fref{fig:resistance}. Note that since the armchair ribbons have a bandgap we average from the valence/conduction band edges for holes/electrons in this specific case. This explains why some of the AC ribbon resistivities are smaller than their corresponding values for the infinite structures. In general, it is seen that the infinite structures are much more electron-hole-symmetric than the ribbons and have lower resistances. This can clearly be seen in the combined mid-right/left-mid resistance,
$$
	R_\text{combined} = \left(\frac{1}{R_\text{MR}} + \frac{1}{R_\text{LM}}\right)^{-1}
$$
which takes into account both sides of the base layer through the junction.

In an experimental realization of the T-junction it is likely that the connection between the two parts consists of many different bonding configurations due to disorder, defects and the specific orientation of the sheets. In such a case, the transmission would be dominated by the most transparent interface and would probably average to some effective resistance over the junction length.

\section{Discussion and Conclusions}
\label{sec:conclusions}

Using \emph{ab initio} calculation methods, we find that three-dimensional graphene T-junctions are energetically feasible when no passivation is present in the connection. For N and O passivation some of the junctions are stable, but unlikely to occur because of their relatively high formation energy. The analysis of junctions created between graphene ribbons and an infinite base sheet using a large ensemble of possible starting configurations, revealed a preferential junction orientation along the armchair direction. Interestingly, this was regardless of whether the ribbons were of zigzag or armchair type.

Utilizing the three-terminal Green's functions transport capabilities of TranSIESTA, we find that the transmissive properties of the junctions depend heavily on the specific connection geometry, but that the transmission from the base graphene sheet to the middle terminal is larger than the one through the base sheet. The contact resistance of various metals end-contacted to graphene encapsulated in hBN have been measured\cite{Wang2013}. The lowest are obtained using Cr ($\sim\!0.1~{\rm k}\Omega\,\mu{\rm m}$), followed by Ti ($\sim\!10~{\rm k}\Omega\,\mu{\rm m}$) and Al, Pd ($\sim\!100~{\rm k}\Omega\,\mu{\rm m}$).

Our results show that not only are such T-junctions feasible, but also in some configurations exhibit superior contact properties. More importantly, the emerging van der Waals technique\cite{geim2013van,Wang2013} provides an ideal platform for creating such edge-plane contact architectures in a controlled and possibly scalable manner. Since a number of techniques to form seamless junctions between carbon nano-structures have been demonstrated experimentally, the T-junction architecture opens for a number of exciting possibilities. While we show that the behavior of the all-graphene junctions is already rich, structurally similar materials that should immediately be possible to join and combine in T-junctions include hexagonal boron nitride and boron-carbon-nitrogen alloys\cite{muchharla2013tunable}, where a mix of boron, carbon and nitrogen atoms allows for tunability of the electrical properties, paving the way for bandgap engineered T-junctions. A T-junction with a single sheet of graphene terminated with a few atomic rows of hBN grown by in-plane heterostructures\cite{liu2013plane} would be a compelling 1D equivalent of a field effect transistor, with the shortest channel imaginable. In fact, the topological similarity with several transistor geometries combined with in-plane heterosynthesis suggests that other 1D analogues of conventional semiconductor components should be possible to realize. Finally, control of the edge chemistry\cite{zhang2012experimentally} opens for a wide range of possibilities in terms of tailoring the structural and electrical properties of such junctions. 

\section*{Acknowledgement}
The authors thank Bjarke S{\o}rensen Jessen, Lene Gammelgaard, and Filippo Pizzocchero for sharing their experimental insight and ideas, and Dr. Gemma Solomon for valuable comments. 
Furthermore, the authors thank the Danish e-Infrastructure Cooperation (DeIC) for providing computer resources, EC Graphene FET Flagship (contract number 604391), and the Lundbeck foundation for support (R95-A10510). The Center for Nanostructured Graphene (CNG) is sponsored by the Danish National Research Foundation (Project DNRF103).

\bibliographystyle{elsarticle-num-names}

\begin{thebibliography}{40}
	\providecommand{\natexlab}[1]{#1}
	\providecommand{\url}[1]{\texttt{#1}}
	\providecommand{\urlprefix}{URL }
	\expandafter\ifx\csname urlstyle\endcsname\relax
	\providecommand{\doi}[1]{doi:\discretionary{}{}{}#1}\else
	\providecommand{\doi}[1]{doi:\discretionary{}{}{}\begingroup
		\urlstyle{rm}\url{#1}\endgroup}\fi
	\providecommand{\bibinfo}[2]{#2}
	
	\bibitem[{Akinwande et~al.(2014)Akinwande, Petrone, and Hone}]{Akinwande2014}
	\bibinfo{author}{D.~Akinwande}, \bibinfo{author}{N.~Petrone},
	\bibinfo{author}{J.~Hone}, \bibinfo{title}{{Two-dimensional flexible
			nanoelectronics}}, \bibinfo{journal}{Nat. Comm.} \bibinfo{volume}{5}
	(\bibinfo{year}{2014}) \bibinfo{pages}{5678}.
	
	\bibitem[{Wang et~al.(2013)Wang, Meric, Huang, Gao, Gao, Tran
		et~al.}]{Wang2013}
	\bibinfo{author}{L.~Wang}, \bibinfo{author}{I.~Meric},
	\bibinfo{author}{P.~Huang}, \bibinfo{author}{Q.~Gao},
	\bibinfo{author}{Y.~Gao}, \bibinfo{author}{H.~Tran}, et~al.,
	\bibinfo{title}{One-Dimensional Electrical Contact to a Two-Dimensional
		Material}, \bibinfo{journal}{Science}
	\bibinfo{volume}{342}~(\bibinfo{number}{6158}) (\bibinfo{year}{2013})
	\bibinfo{pages}{614--617}.
	
	\bibitem[{Novoselov et~al.(2012)Novoselov, Fal'ko, Colombo, Gellert, Schwab,
		and Kim}]{Novoselov2012}
	\bibinfo{author}{K.~Novoselov}, \bibinfo{author}{V.~Fal'ko},
	\bibinfo{author}{L.~Colombo}, \bibinfo{author}{P.~Gellert},
	\bibinfo{author}{M.~Schwab}, \bibinfo{author}{K.~Kim}, \bibinfo{title}{A
		roadmap for graphene}, \bibinfo{journal}{Nature}
	\bibinfo{volume}{490}~(\bibinfo{number}{7419}).
	
	\bibitem[{Geim and Novoselov(2007)}]{Geim2007}
	\bibinfo{author}{A.~Geim}, \bibinfo{author}{K.~Novoselov}, \bibinfo{title}{The
		rise of graphene}, \bibinfo{journal}{Nat. Mater.} \bibinfo{volume}{6}
	(\bibinfo{year}{2007}) \bibinfo{pages}{183--191}.
	
	\bibitem[{Han et~al.(2007)Han, \"{O}zyilmaz, Zhang, and Kim}]{Han2007}
	\bibinfo{author}{M.~Han}, \bibinfo{author}{B.~\"{O}zyilmaz},
	\bibinfo{author}{Y.~Zhang}, \bibinfo{author}{P.~Kim}, \bibinfo{title}{{Energy
			Band-Gap Engineering of Graphene Nanoribbons}}, \bibinfo{journal}{Physical
		Review Letters} \bibinfo{volume}{98}~(\bibinfo{number}{20})
	(\bibinfo{year}{2007}) \bibinfo{pages}{206805}.
	
	\bibitem[{Pedersen et~al.(2008)Pedersen, Flindt, Pedersen, Mortensen, Jauho,
		and Pedersen}]{pedersen_graphene_2008}
	\bibinfo{author}{T.~Pedersen}, \bibinfo{author}{C.~Flindt},
	\bibinfo{author}{J.~Pedersen}, \bibinfo{author}{N.~Mortensen},
	\bibinfo{author}{A.-P. Jauho}, \bibinfo{author}{K.~Pedersen},
	\bibinfo{title}{Graphene Antidot Lattices: Designed Defects and Spin Qubits},
	\bibinfo{journal}{Physical Review Letters}
	\bibinfo{volume}{100}~(\bibinfo{number}{13}) (\bibinfo{year}{2008})
	\bibinfo{pages}{136804}.
	
	\bibitem[{Elias et~al.(2009)Elias, Nair, and Mohiuddin}]{Elias2009}
	\bibinfo{author}{D.~Elias}, \bibinfo{author}{R.~Nair},
	\bibinfo{author}{T.~Mohiuddin}, \bibinfo{title}{Control of graphene's
		properties by reversible hydrogenation: evidence for graphane},
	\bibinfo{journal}{Science} \bibinfo{volume}{323}~(\bibinfo{number}{January})
	(\bibinfo{year}{2009}) \bibinfo{pages}{610--613}.
	
	\bibitem[{Ho et~al.(2014)Ho, Huang, Liao, Zhang, Li, Lai et~al.}]{Ho2014}
	\bibinfo{author}{K.-I. Ho}, \bibinfo{author}{C.-H. Huang},
	\bibinfo{author}{J.-H. Liao}, \bibinfo{author}{W.~Zhang},
	\bibinfo{author}{L.-J. Li}, \bibinfo{author}{C.-S. Lai}, et~al.,
	\bibinfo{title}{Fluorinated graphene as high performance dielectric materials
		and the applications for graphene nanoelectronics},
	\bibinfo{journal}{Scientific reports} \bibinfo{volume}{4}
	(\bibinfo{year}{2014}) \bibinfo{pages}{5893}.
	
	\bibitem[{Sofo et~al.(2007)Sofo, Chaudhari, and Barber}]{Sofo2007}
	\bibinfo{author}{J.~Sofo}, \bibinfo{author}{A.~Chaudhari},
	\bibinfo{author}{G.~Barber}, \bibinfo{title}{Graphane: A two-dimensional
		hydrocarbon}, \bibinfo{journal}{Physical Review B}
	\bibinfo{volume}{75}~(\bibinfo{number}{15}) (\bibinfo{year}{2007})
	\bibinfo{pages}{153401}.
	
	\bibitem[{Oostinga et~al.(2008)Oostinga, Heersche, Liu, Morpurgo, and
		Vandersypen}]{Oostinga2008}
	\bibinfo{author}{J.~Oostinga}, \bibinfo{author}{H.~Heersche},
	\bibinfo{author}{X.~Liu}, \bibinfo{author}{A.~Morpurgo},
	\bibinfo{author}{L.~Vandersypen}, \bibinfo{title}{Gate-induced insulating
		state in bilayer graphene devices}, \bibinfo{journal}{Nat. Mater.}
	\bibinfo{volume}{7}~(\bibinfo{number}{2}) (\bibinfo{year}{2008})
	\bibinfo{pages}{151--157}.
	
	\bibitem[{Ohta et~al.(2006)Ohta, Bostwick, Seyller, Horn, and
		Rotenberg}]{Ohta2006}
	\bibinfo{author}{T.~Ohta}, \bibinfo{author}{A.~Bostwick},
	\bibinfo{author}{T.~Seyller}, \bibinfo{author}{K.~Horn},
	\bibinfo{author}{E.~Rotenberg}, \bibinfo{title}{Controlling the electronic
		structure of bilayer graphene}, \bibinfo{journal}{Science}
	\bibinfo{volume}{313}~(\bibinfo{number}{August}) (\bibinfo{year}{2006})
	\bibinfo{pages}{951--955}.
	
	\bibitem[{Rasmussen et~al.(2013)Rasmussen, Gunst, B{\o}ggild, Jauho, and
		Brandbyge}]{Rasmussen2013}
	\bibinfo{author}{J.~Rasmussen}, \bibinfo{author}{T.~Gunst},
	\bibinfo{author}{P.~B{\o}ggild}, \bibinfo{author}{A.-P. Jauho},
	\bibinfo{author}{M.~Brandbyge}, \bibinfo{title}{Electronic and transport
		properties of kinked graphene}, \bibinfo{journal}{Beilstein J. Nanotechnol.}
	\bibinfo{volume}{4}~(\bibinfo{number}{1}) (\bibinfo{year}{2013})
	\bibinfo{pages}{103--110}.
	
	\bibitem[{Cui et~al.(2015)Cui, Lee, Kim, Arefe, Huang, Lee et~al.}]{Cui2015}
	\bibinfo{author}{X.~Cui}, \bibinfo{author}{G.-H. Lee},
	\bibinfo{author}{Y.~Kim}, \bibinfo{author}{G.~Arefe},
	\bibinfo{author}{P.~Huang}, \bibinfo{author}{C.-H. Lee}, et~al.,
	\bibinfo{title}{Multi-terminal transport measurements of MoS2 using a van der
		Waals heterostructure device platform}, \bibinfo{journal}{Nat. Nanotechnol.}
	\bibinfo{volume}{10}~(\bibinfo{number}{6}) (\bibinfo{year}{2015})
	\bibinfo{pages}{534--540}.
	
	\bibitem[{Fasolino et~al.(2007)Fasolino, Los, and Katsnelson}]{Fasolino2007}
	\bibinfo{author}{A.~Fasolino}, \bibinfo{author}{J.~Los},
	\bibinfo{author}{M.~Katsnelson}, \bibinfo{title}{Intrinsic ripples in
		graphene}, \bibinfo{journal}{Nat. Mater.}
	\bibinfo{volume}{6}~(\bibinfo{number}{11}) (\bibinfo{year}{2007})
	\bibinfo{pages}{858--861}.
	
	\bibitem[{Qi et~al.(2015)Qi, Daniels, Hong, Park, Meunier, Drndić
		et~al.}]{qi2015electronic}
	\bibinfo{author}{Z.~Qi}, \bibinfo{author}{C.~Daniels},
	\bibinfo{author}{S.~Hong}, \bibinfo{author}{Y.~Park},
	\bibinfo{author}{V.~Meunier}, \bibinfo{author}{M.~Drndić}, et~al.,
	\bibinfo{title}{Electronic Transport of Recrystallized Freestanding Graphene
		Nanoribbons}, \bibinfo{journal}{ACS Nano}
	\bibinfo{volume}{9}~(\bibinfo{number}{4}) (\bibinfo{year}{2015})
	\bibinfo{pages}{3510–--3520}.
	
	\bibitem[{Zou et~al.(2012)Zou, Zhang, Xu, Jiang, Tian, Sun
		et~al.}]{zou2012method}
	\bibinfo{author}{R.~Zou}, \bibinfo{author}{Z.~Zhang}, \bibinfo{author}{K.~Xu},
	\bibinfo{author}{L.~Jiang}, \bibinfo{author}{Q.~Tian},
	\bibinfo{author}{Y.~Sun}, et~al., \bibinfo{title}{A method for joining
		individual graphene sheets}, \bibinfo{journal}{Carbon}
	\bibinfo{volume}{50}~(\bibinfo{number}{13}) (\bibinfo{year}{2012})
	\bibinfo{pages}{4965--4972}.
	
	\bibitem[{Wu et~al.(2014)Wu, Zhao, Zhong, Murakawa, and
		Tsukamoto}]{wu2014molecular}
	\bibinfo{author}{X.~Wu}, \bibinfo{author}{H.~Zhao}, \bibinfo{author}{M.~Zhong},
	\bibinfo{author}{H.~Murakawa}, \bibinfo{author}{M.~Tsukamoto},
	\bibinfo{title}{Molecular dynamics simulation of graphene sheets joining
		under ion beam irradiation}, \bibinfo{journal}{Carbon} \bibinfo{volume}{66}
	(\bibinfo{year}{2014}) \bibinfo{pages}{31--38}.
	
	\bibitem[{Wu et~al.(2013)Wu, Zhao, Zhong, Murakawa, and
		Tsukamoto}]{wu2013formation}
	\bibinfo{author}{X.~Wu}, \bibinfo{author}{H.~Zhao}, \bibinfo{author}{M.~Zhong},
	\bibinfo{author}{H.~Murakawa}, \bibinfo{author}{M.~Tsukamoto},
	\bibinfo{title}{The Formation of Molecular Junctions between Graphene
		Sheets}, \bibinfo{journal}{Materials Transactions}
	\bibinfo{volume}{54}~(\bibinfo{number}{6}) (\bibinfo{year}{2013})
	\bibinfo{pages}{940--946}.
	
	\bibitem[{Terrones et~al.(2000)Terrones, Terrones, Banhart, Charlier, and
		Ajayan}]{terrones2000coalescence}
	\bibinfo{author}{M.~Terrones}, \bibinfo{author}{H.~Terrones},
	\bibinfo{author}{F.~Banhart}, \bibinfo{author}{J.-C. Charlier},
	\bibinfo{author}{P.~Ajayan}, \bibinfo{title}{Coalescence of single-walled
		carbon nanotubes}, \bibinfo{journal}{Science}
	\bibinfo{volume}{288}~(\bibinfo{number}{5469}) (\bibinfo{year}{2000})
	\bibinfo{pages}{1226--1229}.
	
	\bibitem[{Zan et~al.(2012)Zan, Ramasse, Bangert, and
		Novoselov}]{zan2012graphene}
	\bibinfo{author}{R.~Zan}, \bibinfo{author}{Q.~Ramasse},
	\bibinfo{author}{U.~Bangert}, \bibinfo{author}{K.~Novoselov},
	\bibinfo{title}{Graphene reknits its holes}, \bibinfo{journal}{Nano Lett.}
	\bibinfo{volume}{12}~(\bibinfo{number}{8}) (\bibinfo{year}{2012})
	\bibinfo{pages}{3936--3940}.
	
	\bibitem[{Yan et~al.(2014)Yan, Peng, Casillas, Lin, Xiang, Zhou
		et~al.}]{Yan2014}
	\bibinfo{author}{Z.~Yan}, \bibinfo{author}{Z.~Peng},
	\bibinfo{author}{G.~Casillas}, \bibinfo{author}{J.~Lin},
	\bibinfo{author}{C.~Xiang}, \bibinfo{author}{H.~Zhou}, et~al.,
	\bibinfo{title}{Rebar graphene}, \bibinfo{journal}{ACS nano}
	\bibinfo{volume}{8}~(\bibinfo{number}{5}) (\bibinfo{year}{2014})
	\bibinfo{pages}{5061--5068}.
	
	\bibitem[{Chen et~al.(2011)Chen, Ren, Gao, Liu, Pei, and Cheng}]{Chen2011b}
	\bibinfo{author}{Z.~Chen}, \bibinfo{author}{W.~Ren}, \bibinfo{author}{L.~Gao},
	\bibinfo{author}{B.~Liu}, \bibinfo{author}{S.~Pei}, \bibinfo{author}{H.-M.
		Cheng}, \bibinfo{title}{Three-dimensional flexible and conductive
		interconnected graphene networks grown by chemical vapour deposition},
	\bibinfo{journal}{Nat. Mater.} \bibinfo{volume}{10}~(\bibinfo{number}{6})
	(\bibinfo{year}{2011}) \bibinfo{pages}{424--428}.
	
	\bibitem[{Lisi et~al.(2011)Lisi, Giorgi, Re, Dikonimos, Giorgi, Salernitano
		et~al.}]{Lisi2011}
	\bibinfo{author}{N.~Lisi}, \bibinfo{author}{R.~Giorgi},
	\bibinfo{author}{M.~Re}, \bibinfo{author}{T.~Dikonimos},
	\bibinfo{author}{L.~Giorgi}, \bibinfo{author}{E.~Salernitano}, et~al.,
	\bibinfo{title}{Carbon nanowall growth on carbon paper by hot filament
		chemical vapour deposition and its microstructure}, \bibinfo{journal}{Carbon
		N.Y.} \bibinfo{volume}{49}~(\bibinfo{number}{6}) (\bibinfo{year}{2011})
	\bibinfo{pages}{2134--2140}.
	
	\bibitem[{Kumar et~al.(2014)Kumar, van~der Laan, Rider, Randeniya, and
		Ostrikov}]{Kumar2014}
	\bibinfo{author}{S.~Kumar}, \bibinfo{author}{T.~van~der Laan},
	\bibinfo{author}{A.~E. Rider}, \bibinfo{author}{L.~Randeniya},
	\bibinfo{author}{K.~K. Ostrikov}, \bibinfo{title}{{Multifunctional
			Three-Dimensional T-Junction Graphene Micro-Wells: Energy-Efficient,
			Plasma-Enabled Growth and Instant Water-Based Transfer for Flexible Device
			Applications}}, \bibinfo{journal}{Advanced Functional Materials}
	\bibinfo{volume}{24}~(\bibinfo{number}{39}) (\bibinfo{year}{2014})
	\bibinfo{pages}{6114--6122}.
	
	\bibitem[{Novaes et~al.(2010)Novaes, Rurali, and Ordej{\'o}n}]{Novaes2010}
	\bibinfo{author}{F.~Novaes}, \bibinfo{author}{R.~Rurali},
	\bibinfo{author}{P.~Ordej{\'o}n}, \bibinfo{title}{Electronic transport
		between graphene layers covalently connected by carbon nanotubes},
	\bibinfo{journal}{ACS Nano} \bibinfo{volume}{4}~(\bibinfo{number}{12})
	(\bibinfo{year}{2010}) \bibinfo{pages}{7596--7602}.
	
	\bibitem[{Brandbyge et~al.(2002)Brandbyge, Mozos, Ordej\'on, Taylor, and
		Stokbro}]{Brandbyge2002}
	\bibinfo{author}{M.~Brandbyge}, \bibinfo{author}{J.~Mozos},
	\bibinfo{author}{P.~Ordej\'on}, \bibinfo{author}{J.~Taylor},
	\bibinfo{author}{K.~Stokbro}, \bibinfo{title}{Density-functional method for
		nonequilibrium electron transport}, \bibinfo{journal}{Phys. Rev. B}
	\bibinfo{volume}{65} (\bibinfo{year}{2002}) \bibinfo{pages}{165401}.
	
	\bibitem[{Papior et~al.(2016)Papior, Gunst, Stradi, and Brandbyge}]{Papior2016}
	\bibinfo{author}{N.~Papior}, \bibinfo{author}{T.~Gunst},
	\bibinfo{author}{D.~Stradi}, \bibinfo{author}{M.~Brandbyge},
	\bibinfo{title}{Manipulating the voltage drop in graphene nanojunctions using
		a gate potential}, \bibinfo{journal}{Phys. Chem. Chem. Phys.}
	\bibinfo{volume}{18} (\bibinfo{year}{2016}) \bibinfo{pages}{1025--1031},
	\doi{\bibinfo{doi}{10.1039/C5CP04613K}}.
	
	\bibitem[{Sol\'{e}r et~al.(2002)Sol\'{e}r, Artacho, Gale, Garc\'{\i}a,
		Junquera, Ordej\'{o}n et~al.}]{Soler}
	\bibinfo{author}{J.~Sol\'{e}r}, \bibinfo{author}{E.~Artacho},
	\bibinfo{author}{J.~Gale}, \bibinfo{author}{A.~Garc\'{\i}a},
	\bibinfo{author}{J.~Junquera}, \bibinfo{author}{P.~Ordej\'{o}n}, et~al.,
	\bibinfo{title}{The SIESTA method for ab initio order-N materials
		simulation}, \bibinfo{journal}{Journal of Physics: Condensed Matter}
	\bibinfo{volume}{14}~(\bibinfo{number}{11}) (\bibinfo{year}{2002})
	\bibinfo{pages}{2745--2779}.
	
	\bibitem[{Kresse and Furthm{\"u}ller(1996)}]{kresse1996efficient}
	\bibinfo{author}{G.~Kresse}, \bibinfo{author}{J.~Furthm{\"u}ller},
	\bibinfo{title}{Efficient iterative schemes for ab initio total-energy
		calculations using a plane-wave basis set}, \bibinfo{journal}{Physical Review
		B} \bibinfo{volume}{54}~(\bibinfo{number}{16}) (\bibinfo{year}{1996})
	\bibinfo{pages}{11169}.
	
	\bibitem[{Perdew et~al.(1996)Perdew, Burke, and Ernzerhof}]{Perdew1996}
	\bibinfo{author}{J.~Perdew}, \bibinfo{author}{K.~Burke},
	\bibinfo{author}{M.~Ernzerhof}, \bibinfo{title}{Generalized Gradient
		Approximation Made Simple}, \bibinfo{journal}{Physical Review Letters}
	\bibinfo{volume}{77}~(\bibinfo{number}{18}) (\bibinfo{year}{1996})
	\bibinfo{pages}{3865--3868}.
	
	\bibitem[{Papior and {et al.}(????)}]{Papior2016a}
	\bibinfo{author}{N.~Papior}, \bibinfo{author}{{et al.}},
	\bibinfo{title}{Unpublished} .
	
	\bibitem[{Saha et~al.(2009)Saha, Lu, Bernholc, and Meunier}]{Saha2009}
	\bibinfo{author}{K.~Saha}, \bibinfo{author}{W.~Lu},
	\bibinfo{author}{J.~Bernholc}, \bibinfo{author}{V.~Meunier},
	\bibinfo{title}{First-principles methodology for quantum transport in
		multiterminal junctions}, \bibinfo{journal}{The Journal of Chemical Physics}
	\bibinfo{volume}{131}~(\bibinfo{number}{16}) (\bibinfo{year}{2009})
	\bibinfo{pages}{164105}.
	
	\bibitem[{Falkenberg and Brandbyge(2015)}]{Falkenberg2015}
	\bibinfo{author}{J.~Falkenberg}, \bibinfo{author}{M.~Brandbyge},
	\bibinfo{title}{{Simple and efficient way of speeding up transmission
			calculations with k-point sampling}}, \bibinfo{journal}{Beilstein J.
		Nanotechnol.} \bibinfo{volume}{6} (\bibinfo{year}{2015})
	\bibinfo{pages}{1603--1608}.
	
	\bibitem[{Felten et~al.(2005)Felten, Bittencourt, Pireaux, {Van Lier}, and
		Charlier}]{Felten2005}
	\bibinfo{author}{A.~Felten}, \bibinfo{author}{C.~Bittencourt},
	\bibinfo{author}{J.~Pireaux}, \bibinfo{author}{G.~{Van Lier}},
	\bibinfo{author}{J.~Charlier}, \bibinfo{title}{{Radio-frequency plasma
			functionalization of carbon nanotubes surface O$_2$, NH$_3$, and CF$_4$
			treatments}}, \bibinfo{journal}{Journal of Applied Physics}
	\bibinfo{volume}{98}~(\bibinfo{number}{2005}) (\bibinfo{year}{2005})
	\bibinfo{pages}{074308}.
	
	\bibitem[{Bittencourt et~al.(2011)Bittencourt, Navio, Nicolay, Ruelle,
		Godfroid, Snyders et~al.}]{Bittencourt2011}
	\bibinfo{author}{C.~Bittencourt}, \bibinfo{author}{C.~Navio},
	\bibinfo{author}{A.~Nicolay}, \bibinfo{author}{B.~Ruelle},
	\bibinfo{author}{T.~Godfroid}, \bibinfo{author}{R.~Snyders}, et~al.,
	\bibinfo{title}{Atomic Oxygen Functionalization of Vertically Aligned Carbon
		Nanotubes}, \bibinfo{journal}{The Journal of Physical Chemistry C}
	\bibinfo{volume}{115} (\bibinfo{year}{2011}) \bibinfo{pages}{20412--20418}.
	
	\bibitem[{Jiang et~al.(2012)Jiang, Turnbull, Lu, Boguslawski, and
		Bernholc}]{Jiang2012}
	\bibinfo{author}{J.~Jiang}, \bibinfo{author}{J.~Turnbull},
	\bibinfo{author}{W.~Lu}, \bibinfo{author}{P.~Boguslawski},
	\bibinfo{author}{J.~Bernholc}, \bibinfo{title}{Theory of nitrogen doping of
		carbon nanoribbons: Edge effects}, \bibinfo{journal}{The Journal of Chemical
		Physics} \bibinfo{volume}{136}~(\bibinfo{number}{1}) (\bibinfo{year}{2012})
	\bibinfo{pages}{014702}.
	
	\bibitem[{Geim and Grigorieva(2013)}]{geim2013van}
	\bibinfo{author}{A.~Geim}, \bibinfo{author}{I.~Grigorieva}, \bibinfo{title}{Van
		der Waals heterostructures}, \bibinfo{journal}{Nature}
	\bibinfo{volume}{499}~(\bibinfo{number}{7459}) (\bibinfo{year}{2013})
	\bibinfo{pages}{419--425}.
	
	\bibitem[{Muchharla et~al.(2013)Muchharla, Pathak, Liu, Song, Jayasekera, Kar
		et~al.}]{muchharla2013tunable}
	\bibinfo{author}{B.~Muchharla}, \bibinfo{author}{A.~Pathak},
	\bibinfo{author}{Z.~Liu}, \bibinfo{author}{L.~Song},
	\bibinfo{author}{T.~Jayasekera}, \bibinfo{author}{S.~Kar}, et~al.,
	\bibinfo{title}{Tunable electronics in large-area atomic layers of
		boron--nitrogen--carbon}, \bibinfo{journal}{Nano Lett.}
	\bibinfo{volume}{13}~(\bibinfo{number}{8}) (\bibinfo{year}{2013})
	\bibinfo{pages}{3476--3481}.
	
	\bibitem[{Liu et~al.(2013)Liu, Ma, Shi, Zhou, Gong, Lei et~al.}]{liu2013plane}
	\bibinfo{author}{Z.~Liu}, \bibinfo{author}{L.~Ma}, \bibinfo{author}{G.~Shi},
	\bibinfo{author}{W.~Zhou}, \bibinfo{author}{Y.~Gong},
	\bibinfo{author}{S.~Lei}, et~al., \bibinfo{title}{In-plane heterostructures
		of graphene and hexagonal boron nitride with controlled domain sizes},
	\bibinfo{journal}{Nat. Nanotechnol.}
	\bibinfo{volume}{8}~(\bibinfo{number}{2}) (\bibinfo{year}{2013})
	\bibinfo{pages}{119--124}.
	
	\bibitem[{Zhang et~al.(2012)Zhang, Yazyev, Feng, Xie, Tao, Chen
		et~al.}]{zhang2012experimentally}
	\bibinfo{author}{X.~Zhang}, \bibinfo{author}{O.~Yazyev},
	\bibinfo{author}{J.~Feng}, \bibinfo{author}{L.~Xie},
	\bibinfo{author}{C.~Tao}, \bibinfo{author}{Y.-C. Chen}, et~al.,
	\bibinfo{title}{Experimentally engineering the edge termination of graphene
		nanoribbons}, \bibinfo{journal}{ACS Nano}
	\bibinfo{volume}{7}~(\bibinfo{number}{1}) (\bibinfo{year}{2012})
	\bibinfo{pages}{198--202}.
	
\end{thebibliography}

\end{document}